# Onset of Instability in Restricted Geometry $^4$He Film Flow

D. H. Liebenberg
49 Starboard Tack Dr.
Salem, SC 29676

## I. Introduction

A brief discussion of $^4$He film flow instability was given[1] as a concluding report to a broader study of thermally driven film flow.[2,3] Since the theoretical work of Feynman[4] considerable effort has been made to identify the properties of quantum turbulence in $^4$He superfluid. The literature is rich with further theoretical and many experimental discussions[5] although a detailed review is outside the scope of the present report. This report has the purpose to examine my earlier research in the light of a remaining problem and new theoretical work in the study of quantum turbulence, namely, the need to examine superfluid turbulence when the normal fluid is immobile and may exhibit a quantum Kelvin-Helmholtz instability[5]. In Section II I review the thermally driven film measurements and discuss the experimental configuration for the restricted geometry measurements. The measurements are discussed in Section III. Discussion is provided in Section IV emphasizing the qualitative similarity of these results with the recent theoretical work of Hiromitsu Takeuchi et al.[5]

## II. Experiments

**Background -** Thermally driven $^4$He saturated superfluid films on the outside of a vertically mounted glass post with nominal 1 cm OD were determined to exhibit dissipation as detected with superconducting granular aluminum thermometers[6] evaporated as a narrow (~1 mm) stripe about half way around the circumference of the post. A heater of Nichrome wire was epoxied to the top of the post for most of the experiments although a copper block with attached heater wire was used for comparison. The post assembly was enclosed in a copper chamber with epoxied polymer windows and immersed in the outer liquid helium bath. The inner helium level between the post thermometer and the bath was adjustable by a externally operated valve. Levels between the post heater and the internal bath could be adjusted from 0 to 15 cm. Temperature control of the outer bath was +/- 5 µK. Leads on the post to the thermometer were evaporated metal sandwiches, copper to provide adhesion to the glass and aluminum, although lead was also tried to provide a superconducting lead. Measurements of the temperature $\Delta T = T_{Al\ thermometer} - T_{bath}$ were recorded as a function of thermal input to the heater dQ/dt. The $\Delta T$ measurements with the present apparatus were stable over times of more than 30 minutes and the response time to changes was within the chart recorder response, ~1 sec. The relations between input heater power and flow velocity and between the $\Delta T$ and dissipation were shown in Ref. 2,3. The main result of those data was that the magnitude of the dissipation was related to the superfluid velocity by the Iordanskii-Langer-Fisher fluctuation theory[7]. Further, the results of thermally driven flow were compared[8] with the dissipation from the clever capillary potential probe measurement technique of Keller and Hammel[9]. Again good agreement was found for the dissipation fluctuation picture of the thermally driven film; the $\Delta T$ measurements and potential probe measurements overlapped. Additionally, the potential probe measurements extended to lower dissipation levels.

At higher power levels a step in the measurements was observed[3]; a plateau of nearly constant ΔT was found as the power was increased and then an additional increase in ΔT until saturation occurred at a still higher power level. For these higher power measurements, after the power level was established the ΔT remained constant indicating that there was no burn-off of the film. For isothermal gravitational flow it had been established that no change in film thickness occurred attributable to dissipation[10].

**Restricted Geometry Experiment -** With the earlier experiments as backdrop a new experiment was designed and preliminary results obtained before programmatic requirements closed the project. A glass post of 12 mm diameter was notched between the bath and the heater. The notch was ~ 0.25 mm wide and tapered to 0.025 mm at the bottom of the notch that was 1.5 mm deep. The cross-section area of the helium film was reduced by about 12 % in this region. This reduction would confine the dissipation to this small region and increase the gradient of the dissipation. A superconducting granular aluminum thermometer (GAT) was evaporated into the notch and leads prepared as before so that ΔT measurements were made at the region of dissipation. These leads were evaporated copper down the glass post and connected to copper wires through superfluid tight seals in the copper base plate. Calibration of the GAT was carried out by determining the resistance change with temperature as measured with an oil manometer or a Texas Instruments pressure gauge and standard tables. The sensitivity of the GAT was 60 Ω/K in the temperature range 1.35 - 1.45 K, reduced from more typical values, ~ 2 x $10^4$ Ω/K, due apparently to reduced granulation.[6] The heater of Nichrome wire was installed some distance below the top of the post and about 8.25 mm above the notch. The glass post was positioned in the closed copper cylinder that was provided with polymer windows and sealed with Stycast®. Helium levels were measured with a cathetometer. This assembly was submerged in an external liquid helium bath maintained at a selected temperature with a controller to +/- 5 μK. The temperature was measured with a manometer and a Texas Instruments Pressure Gauge. Liquid helium was introduced into the copper chamber with the glass post via a superfluid tight valve connecting to the external bath. A Keithley® model 148 nanovoltmeter was connected to the voltage leads of the aluminum thermometer operated in a four wire configuration. The output was recorded on a strip chart recorder.

### III. Results

The observed growth of dissipation at low flow velocities (low heater power) was similar to that observed for the constant diameter posts and was characterized by the fluctuation-dissipation theory[2]. At higher heater power the onset of saturation was observed as before[3]. However, well prior to saturation an instability occurred as shown in Fig. 1. The chart recording rather than a draftsperson's rendition is shown for a part of one of two different helium runs. The recorder is running at 1 in./min with 1 in. being the major division lines. Both at lower and higher power levels there was significantly less instability as denoted by the relatively constant voltage recorded. The nearly constant voltage at each power level setting indicates that there is no burn off of the film or temporal interaction with the surrounding vapor. As seen in Fig. 1 a constant power level and response was held for 30 to 60 sec. Even at the saturation power level there is a constant temperature over this duration. Nor is there voltage drift for the power settings where instabilities or oscillations are observed. The change of heater power between the initiation and termination of the oscillations is only a small percentage of the saturation power so these oscillations are apparently confined to a narrow film flow velocity range. The oscillation period on the chart was of the order of the response time (1 s) and so does not provide much insight into the phenomena. Nor does the amplitude properly represent the voltage fluctuations. However, if the amplitude of the fluctuations at the reduced sensitivity is considered, the maximum amplitude represents a 20 mK temperature swing. This would represent a superfluid density decrease of about 5%[12] The measurements in Fig. 1 were made with the bath level at approximately 27.5 mm below the notch as determined by a cathetometer measurement. Using earlier results for film thickness[11] a value of about 22 nm is determined for the static film. A series of bath level changes were made and oscillations measured. At the temperature of the fluid reservoir, 1.37 K, the normal fluid density represents about 6.5% of the total density and in the film is immobile[12]. A second experimental run gave similar oscillation results with the bath level at 85 mm below the notch and at about the same reservoir temperature. The oscillations onset at a lower heater input power as would be consistent with the reduced static film thickness of about 16 nm.

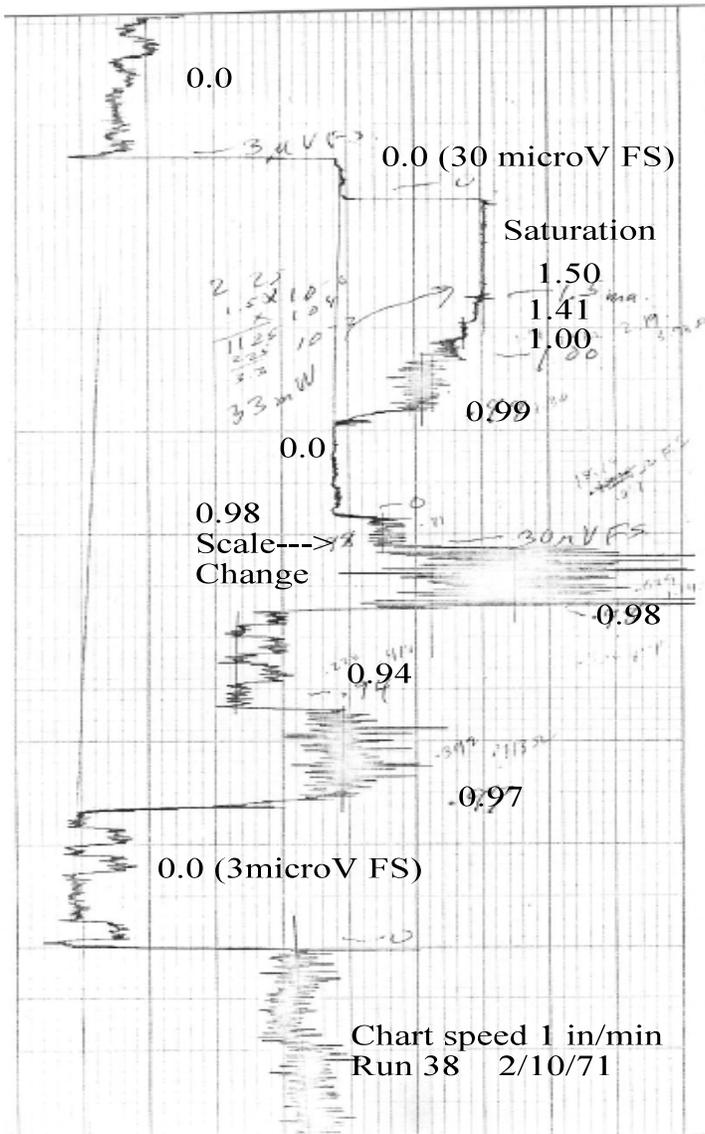

Fig. 1 Copy of part of the chart recording to show the onset of oscillations that have a frequency greater than the chart recorder bandwidth. Relative power levels are marked as is a change in sensitivity.

### IV. Discussion

To assure that the GAT was operating as would be expected for a normal[2] dissipation measurement a series of $\Delta T$ vs height H of the notch above the liquid level was performed. At constant power level of 10.1 mW the $\Delta T$ scaled as $\sim H^{1/3}$ equivalent to a thickness variation of $\sim H^{-1/3}$. This is the thickness variation expected for heights above the liquid greater than about 1 cm. In addition several onset power levels (OP) were obtained versus H and again the measurements were well described as OP $\sim H^{1/3}$. These results are consistent with thickness vs H measurements of the static film and suggest that no change in film thickness is attributable to the dissipation even at these higher power levels; in agreement with the results of Ref. 10.

This result, together with the presumption, Ref. 12, that the normal fluid fraction in the film is immobile suggests that the onset of oscillating instability of the thermally driven helium film dissipation as determined from ΔT is likely related to quantum turbulence under conditions substantially different than those of vibrating wires or moving grids as recently reviewed in Ref. 5. The recent theory of Hiromitsu Takenuchi et al. (TSKST) of Ref. 5 is of particular interest since a flat interface between the superfluid and non-mobile normal component is shown to lead to the development of sawtooth waves. In the present case the difference in superfluid to normal fluid ratio would result in a temperature variation such as is detected in this experiment. Further, at larger relative flow velocities these waves tend to smooth again which would be consistent with the cutoff of oscillations seen at the higher driving force levels of the present results. At present I can not rule out that the lack of continuing oscillations at higher power levels might be the result of the sensor/recorder response to higher frequency oscillations. The lack of drift in ΔT for 10's of seconds after the onset suggests a stability of the film in this mode. The amplitude of the oscillations remains within the nominal linearity of the GAT as determined from the calibration. The film velocity remains below the roton critical velocity of Ref. 4.

Clearly these preliminary experiments can be repeated now, more than 36 years later, with improved instrumentation and controls. The thin film aluminum temperature sensors do offer microKelvin sensitivity. Bath control may be improved but improvement is probably not important. Baffling could be used to further control vapor interactions although the 60 mm diameter copper cylinder was designed to minimize wall and vapor interactions with the 12 mm diameter glass post. I hope these reported results will stimulate further work on this approach to the study of quantum turbulence in liquid $^4$He under conditions of an immobilized normal fluid fraction. In view of the theory suggested by TSKST such experiments might provide more quantitative comparison.

**Acknowledgments**
This thesis research was performed at the Los Alamos National Laboratory and I thank Drs. R. Fowler (deceased), L. Allen (deceased), E. F. Hammel, Wm. Keller, and L. Campbell, and also the glass shop for contributions of equipment, support, and advice for this research. Prof. J. R. Dillinger (deceased), University of Wisconsin, stimulated my interest in helium superfluid research and was my thesis advisor. I also thank Dr. Thomas Jones (deceased) National Science Foundation Division Director who encouraged this effort.